\documentclass{article}

\usepackage[preprint]{nips_2018}

\usepackage[utf8]{inputenc}	
\usepackage[T1]{fontenc}		
\usepackage{hyperref}		
\usepackage{url}			
\usepackage{booktabs}		
\usepackage{amsfonts}		
\usepackage{nicefrac}		
\usepackage{microtype}		
\usepackage{graphicx}
\usepackage{listings}
\usepackage{color}
\usepackage{todonotes}
\usepackage{subfig}
\usepackage{wrapfig}
\usepackage[font=small,labelfont=bf]{caption}

\usepackage{float}
\floatstyle{boxed}
\restylefloat{figure}
\restylefloat{table}

\newcommand{\server}[1]{\ensuremath{ S_{#1} }}

\newcommand{\private}[1] {\ensuremath{ \langle #1 \rangle }}
\newcommand{\masked}[1] {\ensuremath{ \langle \! \langle #1 \rangle \! \rangle }}
\newcommand{\share}[2]{\ensuremath{ {#2}_{#1} }}

\title{Private Machine Learning in TensorFlow \\ using Secure Computation}

\author{
  Morten Dahl
  \And
  Jason Mancuso
  \And
  Yann Dupis
  \And
  Ben DeCoste
  \AND
  Morgan Giraud
  \And
  Ian Livingstone
  \And
  Justin Patriquin
  \And
  Gavin Uhma
  \AND
  \textit{Dropout Labs}
}

\begin{document}

\maketitle
\vspace{-15px}
\begin{abstract}
We present a framework for experimenting with secure multi-party computation directly in TensorFlow. By doing so we benefit from several properties valuable to both researchers and practitioners, including tight integration with ordinary machine learning processes, existing optimizations for distributed computation in TensorFlow, high-level abstractions for expressing complex algorithms and protocols, and an expanded set of familiar tooling. We give an open source implementation of a state-of-the-art protocol and report on concrete benchmarks using typical models from private machine learning.
\end{abstract}
\vspace{-10px}

\section{Introduction}
Several fields come together in private machine learning: cryptography, machine learning, distributed systems, and high-performance computing. As a result, researchers and practitioners developing scalable solutions may find themselves faced with a task requiring many diverse skill sets and expertise.

Adapting machine learning models in a way that allows for privacy-preserving prediction and training is complex and non-trivial on its own, and often requires deeper insight into cryptography or machine learning than experts from either field have about the other.
For instance, one may want to mix secure computation protocols with complementary properties and experiment with various cryptographic optimizations such as vectorization and other specializations. Simultaneously, one may want to find variants of machine learning components that give sufficient accuracy yet utilize the fact that some secure operations are significantly faster than others. Thus, modularity and extensibility are crucial for effective experimentation.

One may additionally be faced with a large implementation challenge, potentially having to stitch together several otherwise independent systems or re-implement complex methods from both fields. One must not only optimize local computations by taking advantage of diverse multi-core architectures, but also the distributed processes inherent in secure computation, including how best to orchestrate the joint execution on multiple machines and minimize the overhead of sending large amounts of data across the network. Building all of this at the right level of abstraction can be overwhelming, and often comes at the cost of extensibility and code readability. This makes experimentation harder, potentially harming accessibility and correctness.

Finally, for even small-scale solutions it is highly valuable to have access to tools for visual inspection, debugging, and profiling such as TensorBoard in order to identify issues and bottlenecks in both the protocol and the machine learning model. Lack of such tools represents an additional barrier to entry.

In this paper we illustrate the benefits of embedding protocols directly into TensorFlow in order to show that it can serve as a platform for easily experimenting with secure computation for private machine learning.

\setlength{\intextsep}{0pt}%
\setlength{\columnsep}{15pt}%
\begin{wrapfigure}{R}{0.38\textwidth}
\centering
\lstset{
    language=Python,
    morekeywords={with},
    basicstyle=\fontsize{6}{7}\selectfont\ttfamily,
    keywordstyle=\color{blue}\ttfamily,
    stringstyle=\color{red}\ttfamily,
    commentstyle=\color{darkgreen}\ttfamily,
    breaklines=false,
}
\vspace{-5px}
\begin{lstlisting}
import tensorflow as tf
import tf_encrypted as tfe

@tfe.private_input('model-owner')
def send_model_weights() -> tf.Tensor:
    ...

@tfe.private_input('prediction-client')
def send_input() -> tf.Tensor:
    ...

@tfe.private_output('prediction-client')
def receive_output(y: tf.Tensor):
    return tf.Print([], [y])

w = send_model_weights()
x = send_input()
y = tfe.sigmoid(tfe.matmul(x, w))
op = receive_output(y)

with tfe.Session() as sess:
    sess.run(op)
\end{lstlisting}
\caption{Example using \texttt{tf-encrypted} for private prediction,
with the prediction input known only in plaintext by the client and the model weights only by the owner.}
\label{fig:pred}
\vspace{-5px}
\end{wrapfigure}

\subsection{Contributions}


We present \texttt{tf-encrypted}, an open source library built on top of TensorFlow with the aim of making private machine learning more accessible to researchers and practitioners coming from either cryptography or machine learning, and without the need to be an expert in distributed systems or high-performance computing. A small example is given in Figure~\ref{fig:pred}, where a logistic regression model is used for making private predictions. Note that the methods processing input and output specify on which party they should run, and that usual TensorFlow operations may be used within to perform local plaintext operations.

To this end we adapt and implement a state-of-the-art secure computation protocol for tensor oriented applications (Section~\ref{sec:pond}). We report on benchmarks using common models from the literature (Section~\ref{sec:experiments}) and highlight additional properties of this approach that we find of value:
\textbf{usability:} by leveraging TensorFlow we obtain a familiar and comprehensive platform for building scalable solutions;
\textbf{integration:} by reducing all secure computations to TensorFlow graphs, it becomes straight-forward to mix these with ordinary computations for hybrid approaches;
\textbf{extensibility:}
using TensorFlow's high-level abstractions makes it easier to experiment with and develop new secure protocols on top of optimized primitives while maintaining code readability;
\textbf{performance:} we reach high runtime efficiency without sacrificing other properties via TensorFlow's distributed execution engine heavily optimized for networking, parallel execution, and scalability;
\textbf{benchmarking:} combining all of the above we obtain a common framework for comparable private machine learning.

\subsection{Related Work}

Several freely available implementations of secure computation protocols exist, including those of~\cite{SCALEMAMBA, FRESCO, DSZ15, OblivC}, yet all of these are standalone frameworks that do not provide integration with existing machine learning platforms, and arguably aim more at general purpose secure computation than our focus on private machine learning. As a result, users are faced with a lower level interface and may have to implement basic machine learning components. We are furthermore not aware of any development tools for these frameworks outside of general purpose debuggers.
To the best of our knowledge no machine learning support tools exist for these platforms.

The works of~\cite{SecureML, Gazelle, SecureNN, ABY3} focus on adapting secure computation protocols to private machine learning, in some cases using similar optimizations as the protocol presented in Section~\ref{sec:pond}. However, to the best of our knowledge none of these have openly available implementations and hence require a significant investment from anyone wanting to apply them. To a large extent our aim is to provide a common platform based on TensorFlow for implementing and experimenting with protocols such as these.
The concurrent work of~\cite{PySyft} takes a somewhat similar approach yet currently focuses on PyTorch as opposed to TensorFlow.

Another line of work has focused on using differential privacy for privacy-preserving machine learning~\cite{PAEGT17, PSMRTE18}, with reference implementations available in TensorFlow. These works remain orthogonal to our approach and do not employ any form of secure computation.

\section{Secure Computations in TensorFlow}
\label{sec:pond}

TensorFlow as described by~\cite{TensorFlow16} is among the leading frameworks for constructing and deploying machine learning models, offering an optimized engine for executing local and distributed computations as well as a high-level interface for expressing these. Importantly, the latter abstracts away lower-level operations such as networking while remaining powerful enough to succinctly express computations with operations and tensors pinned to specified machines (see for example Figure~\ref{fig:mul}). This combines for a powerful platform for giving efficient implementations of distributed computations, including complex secure computations where it is crucial for privacy that some data remain known only on select machines.

Moreover, the link between TensorFlow's engine and high-level interface takes the form of stateful dataflow graphs, with nodes for performing tensor operations in the distributed setting. This means that the engine can not only take advantage of optimizations such as lazy evaluation and multi-core processing tailored for the specific runtime environment, but also optimize the graph itself and chose a node execution order based on both static and runtime information (see e.g.~\cite{TensorFlow15}). For instance, we observe that large network transfers from Beaver triple generation are often automatically batched and moved to the beginning of the execution.

As an example of a secure computation protocol implemented in TensorFlow, we here outline our variant of the well-known SPDZ protocol by~\cite{SPDZ12} with two servers $\server{0}$ and $\server{1}$ that is used for benchmarking in Section~\ref{sec:experiments}. The protocol is vectorized to improve performance of applications relying heavily on tensor operations and we use generalized triples produced by an independent third server\footnote{This server is essentially the crypto producer as used in e.g.~\cite{SecureML} that can run entirely offline as long as the function to compute is known, which can be fully determined at compile time due to TensorFlow's use of static dataflow graphs.} to avoid sending redundant data when possible. Any number of input providers and output receivers are supported, holding e.g. training data or prediction inputs.  We currently ensure passive (honest-but-curious) security under a single corruption and rely on two cryptographic primitives, namely 
additive secret sharing and secure channels between all players.

\setlength{\intextsep}{0pt}%
\setlength{\columnsep}{15pt}%
\begin{wrapfigure}{R}{0.42\textwidth}
\centering
\lstset{
    language=Python,
    morekeywords={with},
    basicstyle=\fontsize{6}{7}\selectfont\ttfamily,
    keywordstyle=\color{blue}\ttfamily,
    stringstyle=\color{red}\ttfamily,
    commentstyle=\color{darkgreen}\ttfamily,
    breaklines=false,
}
\vspace{-5px}
\begin{lstlisting}
def mul(x: MaskedTensor, y: MaskedTensor):
    a, a0, a1, alpha0, alpha1 = x.unwrapped
    b, b0, b1, beta0, beta1 = y.unwrapped

    with tf.name_scope('mul'):

        with tf.device(crypto_producer):
            ab = a * b
            ab0, ab1 = share(ab)

        with tf.device(server_0):
            z0 = ab0 + (a0 * beta0) \
                 + (alpha0 * b0) \
                 + (alpha0 * beta)

        with tf.device(server_1):
            z1 = ab1 + (a1 * beta1) \
                 + (alpha1 * b1)

    return PrivateTensor(z0, z1)
\end{lstlisting}
\caption{Secure multiplication implemented in TensorFlow: \texttt{tf.device} pin data and operations to specific machines to multiply two masked tensors, and takes care of implicitly adding nodes for transmitting \texttt{ab0} and \texttt{ab1}.}
\label{fig:mul}
\vspace{-5px}
\end{wrapfigure}

Following typical practice we use a fixed point encoding for the floating point numbers commonly used in machine learning, i.e. we scale by a fixed factor and treat the result as an integer. To represent these we support both a fixed \texttt{int64} and a CRT-based \texttt{int100} tensor type, the former offering higher performance and the latter higher precision. To maintain precision after multiplications we implement both the conservative truncation protocol of~\cite{CS10} requiring one round of communication, as well as the optimistic non-interactive protocol of~\cite{SecureML} that may fail with a small probability.

As in other SPDZ variants, \textit{private tensors} $\private{x}$ are secret shared into two tensors $\share{0}{x}$ and $\share{1}{x}$, held by server $\server{0}$ and $\server{1}$ respectively, such that $x = \share{0}{x} + \share{1}{x}$ yet either share on its own reveals nothing about $x$. However, unlike other variants, we also rely on a \textit{masked tensor} $\masked{x}$ which in addition to $\share{0}{x}$ and $\share{1}{x}$ also includes a random tensor $a$ held by $\server{2}$, shares $\share{0}{a}$ and $\share{1}{a}$ of it held by $\server{0}$ and $\server{1}$, and an $\alpha$ held by both $\server{0}$ and $\server{1}$ such that $\alpha = x - a$. While this is simply an explicit representation of the intermediate state of a SPDZ multiplication, having it in this form allows us to easily extend and generalize triples in order to reduce computation and networking.
Converting a tensor from private to masked takes one round of \emph{offline} communication where $a$ is sampled by $\server{2}$ and shares $\share{0}{a}$ and $\share{1}{a}$ of it are sent to $\server{0}$ and $\server{1}$, and one round of \emph{online} communication where $\share{i}{x} - \share{i}{a}$ is sent by $\server{i}$ to $\server{1-i}$ for $i \in \{0, 1\}$ and $(\share{0}{x} - \share{0}{a}) + (\share{1}{x} - \share{1}{a}) = \alpha$ is computed by both.


Secure computation on tensors proceed as in other SPDZ variants, albeit with some operations first converting private tensors into masked tensors. For instance, 
multiplication is only implemented for masked tensors as $\private{z} = \mathsf{mul}(\masked{x}, \masked{y})$ with $\share{i}{z} = i \alpha^x \alpha^y + \alpha^x \share{i}{a}^y + \share{i}{a}^x \alpha^y + \share{i}{a^x a^y}$ and where $a^x a^y$ is computed and shared by $\server{2}$ (see Figure~\ref{fig:mul}).
Note that the result of a multiplication is a private tensor, meaning $z$ will have to be masked before it can be used as input to another multiplication; however, both $\masked{x}$ and $\masked{y}$ can readily be used again as is, meaning every tensor only needs to be masked once.
While special triples for squaring are no longer needed with explicit masking, we can still optimize e.g. matrix multiplications and convolutions via specialized triples and avoid reducing everything to (scalar) multiplications, in turn reducing networking further. Finally, operations such as transposing and stacking are done locally without interaction by letting all three servers operate on the values associated with private and masked tensors.

%

\section{Experiments}
\label{sec:experiments}

We benchmark the protocol of Section~\ref{sec:pond} for private inference on the typical MNIST handwritten digit classification task. Using TensorFlow we train each of the neural networks\footnote{These are variants of models studied in \cite{SecureML, Gazelle, SecureNN}. We use polynomials interpolated to fit ReLU on the interval $[-3,3]$ for activation functions and perform the final argmax and softmax on logits in plaintext to avoid computing these securely.} in Figure~\ref{fig:networks} on the plaintext training set and then run private inferences on the remaining test set, keeping both prediction input and model weights private. We perform all experiments on the Google Cloud Platform using instances in the same region (\texttt{us-east1}) and with 36 vCPUs/60 GB memory each. We note that TensorBoard was invaluable in this process for picking the right approximation intervals for activation functions and inspecting overall correctness, and that performing all operations in TensorFlow simplified the process of model handling significantly.

\begin{wrapfigure}{L}{0.55\textwidth}
\small
\centering
\begin{tabular}{ccc}
\textbf{Network A}
& \textbf{Network B}
& \textbf{Network C}
\\ \hline
\begin{tabular}[c]{@{}l@{}}FC (784, 128) \\ BatchNorm\\ ReLU (approx) \\ FC (128, 128) \\ BatchNorm\\ ReLU (approx) \\ FC (128,  10) \end{tabular} & \begin{tabular}[c]{@{}l@{}}Conv (5, 16, 1, 1) \\ BatchNorm\\ ReLU (approx) \\ AvgPool (2) \\ Conv (5, 16, 1,  1) \\ BatchNorm \\ ReLU (approx) \\ AvgPool (2) \\ FC (256, 100) \\ BatchNorm\\ ReLU (approx) \\ FC (100, 10) \end{tabular} & \begin{tabular}[c]{@{}l@{}}Conv (5, 20, 1, 1) \\ BatchNorm\\ ReLU (approx) \\ AvgPool (2) \\ Conv (5, 50, 1, 1) \\ BatchNorm\\ ReLU (approx) \\ AvgPool (2) \\ FC (800, 500) \\ BatchNorm\\ ReLU (approx) \\ FC (500, 10) \end{tabular}
\end{tabular}

\caption{Neural network architectures. The convolutional layers are denoted by $(\textit{field size}, \textit{channels}, \textit{stride}, \textit{padding})$ and average pooling layers by $\textit{window size}$}
\label{fig:networks}
\end{wrapfigure}

The left part of Table~\ref{tab:perf} summarizes the combined offline and online runtime averaged over 100 inferences. As seen we get reasonable performance that may already be adequate for concrete applications. The fact that \texttt{int100} also achieves good performance suggests that we are not limited by the lower precision of \texttt{int64} when looking at larger models. Additional experiments furthermore indicate sub-linearly scaling with respect to typical batch sizes, leading to an interesting trade-off between latency and through-put; for instance, Network C with batch sizes 1, 10, and 100 take respectively 124, 182, and 541ms.

We also compare our accuracy over the entire testing set against plaintext TensorFlow. As seen in the right part of Table~\ref{tab:perf} we obtain almost identical accuracy using both \texttt{int64} and \texttt{int100}, with indications that the latter may give a slightly better output distribution according to the mean KL divergence, in turn potentially justifying its higher runtime cost. We note that these models achieve good performance despite using approximations, further underlining the importance of being able to adapt models to the encrypted setting.

Based on the above we conjecture that our approach scales to larger models, and is at least within an order of magnitude of related work in terms of runtime performance (at what we believe to be a lower implementation cost). We defer proper exploration of both topics to the full version of this paper.


\begin{table}
\small
\centering
\begin{tabular}{l|r|r|r|r|r|r|l|l|l}
& \multicolumn{2}{c|}{Runtime average}
& \multicolumn{2}{c|}{Runtime deviation}
& \multicolumn{3}{c|}{Accuracy}
& \multicolumn{2}{c}{KL divergence} \\
& \multicolumn{1}{c}{\texttt{int64}}
& \multicolumn{1}{c|}{\texttt{int100}} 
& \multicolumn{1}{c}{\texttt{int64}} 
& \multicolumn{1}{c|}{\texttt{int100}}
& \multicolumn{1}{c}{TF}
& \multicolumn{1}{c}{\texttt{int64}}
& \multicolumn{1}{c|}{\texttt{int100}}
& \multicolumn{1}{c}{\texttt{int64}}
& \multicolumn{1}{c}{\texttt{int100}}\\ \hline
\multicolumn{1}{c|}{A} &  14ms & 138ms & 3.8ms & 61ms & 97.35\% & 97.18\% & 97.26\% & 0.0065 & 0.0064 \\
\multicolumn{1}{c|}{B} & 126ms & 189ms & 115ms & 94ms & 99.26\% & 99.00\% & 98.93\% & 0.2086  & 0.0311 \\
\multicolumn{1}{c|}{C} & 124ms & 211ms &  93ms & 60ms & 99.44\% & 99.41\% & 98.54\% & 0.2311 & 0.1045
\end{tabular}
\caption{Runtime benchmarks with \texttt{int64} and \texttt{int100}.}
\label{tab:perf}
\end{table}


 


\section{Conclusion}

We have proposed an open source framework for experimenting with secure computation in TensorFlow, and illustrated how implementation of such protocols can be easily expressed using high level abstractions (Figure~\ref{fig:mul}). This additionally allows private machine learning to be expressed in an interface similar to ordinary TensorFlow (Figure~\ref{fig:pred}) while maintaining good performance.

In the full version of this paper we elaborate on these results, and present a modular extension of the concrete protocol presented here that adds features from~\cite{SecureNN} in order to compute exact ReLU and MaxPooling.

\bibliographystyle{plainnat}
\bibliography{references}

\end{document}